Inelastically generated solitons and anti-solitons in the perturbed KdV equation

Yair Zarmi

Jacob Blaustein Institutes for Desert Research
& Physics Department
Ben-Gurion University of the Negev
Midreshet Ben-Gurion, 84990
Israel

Abstract

Under the effect of common perturbations, the multiple-soliton solution of the KdV equation is transformed into a sum of an elastic and a first-order inelastic component. The elastic component is a perturbation series, identical in structure to the perturbed single-soliton solution. It preserves the soliton-scattering picture. The inelastic component is generated by perturbation terms that represent coupling between KdV solitons and inelastically generated soliton-anti-soliton waves. It asymptotes into solitons and anti-solitons, that evolve along the characteristic lines of the KdV solitons. This is demonstrated in the two-soliton case.

A common form of the perturbed KdV equation is [1-6]:

$$\begin{aligned}
w_t &= 6ww_1 + w_3 \\
&+ \varepsilon\left(30\alpha_1 w^2 w_1 + 10\alpha_2 ww_3 + 20\alpha_3 w_1 w_2 + \alpha_4 w_5\right) \\
&+ \varepsilon^2 \begin{pmatrix} 140\beta_1 w^3 w_1 + 70\beta_2 w^2 w_3 + 280\beta_3 ww_1 w_2 + 14\beta_4 ww_5 + 70\beta_5 w_1^3 \\ + 42\beta_6 w_1 w_4 + 70\beta_7 w_2 w_3 + \beta_8 w_7 \end{pmatrix} \\
&+ \varepsilon^3 \begin{pmatrix} 630\gamma_1 w^4 w_1 + 1260\gamma_2 w(w_1)^3 + 2520\gamma_3 w^2 w_1 w_2 + 1302\gamma_4 w_1 (w_2)^2 \\ + 420\gamma_5 w^3 w_3 + 966\gamma_6 (w_1)^2 w_3 + 1260\gamma_7 ww_2 w_3 + 756\gamma_8 ww_1 w_4 \\ + 252\gamma_9 w_3 w_4 + 126\gamma_{10} w^2 w_5 + 168\gamma_{11} w_2 w_5 + 72\gamma_{12} w_1 w_6 + 18\gamma_{13} ww_7 + \gamma_{14} w_9 \end{pmatrix} \\
&+ O(\varepsilon^4) \qquad\qquad \left(|\varepsilon| \ll 1, \; w_p \equiv \partial_x^p w\right)
\end{aligned} \quad (1)$$

The unspecified coefficients depend on the dynamical system for which Eq. (1) is derived. For example, small-amplitude solutions of surface waves on a shallow-water-layer over a horizontal plane [7] and of the ion acoustic wave equations in Plasma Physics [8, 9] are approximately described by Eq. (1) [10].

Eq. (1) is asymptotically integrable through $O(\varepsilon)$ [1]. Namely, if $w$ is expanded in powers of $\varepsilon$,

$$w(t,x) = u(t,x) + \varepsilon u^{(1)}(t,x) + \varepsilon^2 u^{(2)}(t,x) + \varepsilon^3 u^{(3)}(t,x) + O(\varepsilon^4) , \qquad (2)$$

then $u^{(1)}$ can be expressed as a differential polynomial in the zero-order approximation, $u$,

$$u^{(1)} = a_1 u_2 + a_2 u^2 + a_3 u_1 q \qquad \left(q(t,x) = \int_{-\infty}^{x} u(t,x)dx\right) , \qquad (3)$$

and $u$ is determined by a Normal Form that is integrable through $O(\varepsilon)$,

$$u_t = 6uu_1 + u_3 + \varepsilon\,\alpha_4\left(30u^2 u_1 + 10uu_3 + 20u_1 u_2 + u_5\right) + O(\varepsilon^2) . \qquad (4)$$

Eq. (4) has the same soliton solutions as the KdV equation. Denoting the wave number of the $i^{th}$ soliton by $k_i$, the velocity of each soliton in a multiple-soliton solution is updated according to:

$$v_i = 4k_i^2\left(1 + \varepsilon\alpha_4 4k_i^2 + O(\varepsilon^2)\right) . \tag{5}$$

Let us now turn to the first-order correction, $u^{(1)}$, of Eq. (3). When $u$ is a single-soliton solution, the freedom in the equations allows for the determination of two of the coefficients only [1-6]:

$$a_1 = a_3 - \tfrac{5}{2}\alpha_1 + \tfrac{10}{3}\alpha_2 + \tfrac{5}{3}\alpha_3 - \tfrac{5}{2}\alpha_4 \quad , \quad a_2 = a_3 - 5\alpha_1 + 5a_2 . \tag{6}$$

For the general solution, no such freedom exists, and $a_3$ obtains the value [1-6]:

$$a_3 = -\tfrac{10}{3}\alpha_2 + \tfrac{10}{3}\alpha_4 . \tag{7}$$

When $u(t,x)$ is a multiple-soliton solution, it displays a simple scattering picture. Far from the soliton-collision region (a finite domain around the origin in the $x$-$t$ plane), each soliton maintains its functional form before and after the collision. It may be affected by the presence of the other solitons at most through a trivial phase shift. However, the scattering picture is spoiled in $O(\varepsilon)$, by the term that is proportional to the non-local quantity, $q(t,x)$, in Eq. (3). (Although this term is localized around the characteristic line of each soliton, as it is proportional to $u_x$.) This has been demonstrated [11] in the case of the two-soliton solution, given by [12]

$$u(t,x) = 2\partial_x^2 \ln\left\{1 + g_1(t,x) + g_2(t,x) + \left(\frac{k_1 - k_2}{k_1 + k_2}\right)^2 g_1(t,x)g_2(t,x)\right\}$$
$$\left(g_i(t,x) = \exp[2k_i(x + v_i t)]\right) \tag{7}$$

Here $v_i$ are given by Eq. (5). Assuming $k_1 < k_2$, $q(t,x)$ asymptotes to the following values:

$$q(t,x) \to \begin{cases} \overbrace{\begin{matrix} 0 & , & x \ll -4k_2^2 t \\ 2k_1 & , & -4k_2^2 t \ll x \ll -4k_1^2 t \\ 2(k_1 + k_2) & , & -4k_1^2 \ll x \end{matrix}}^{t<0} & \overbrace{\begin{matrix} 0 & , & x \ll -4k_1^2 t \\ 2k_2 & , & -4k_1^2 t \ll x \ll -4k_2^2 t \\ 2(k_1 + k_2) & , & -4k_2^2 \ll x \end{matrix}}^{t>0} \end{cases} . \tag{8}$$

Consequently, the correction to the amplitude of soliton no. 1 (no. 2) is affected by $k_2$ ($k_1$) for $t < 0$ ($t > 0$), respectively. Integrability is achievable in first order, but the scattering picture is lost.

To preserve the scattering picture in $O(\varepsilon)$, one must eliminate the term proportional to $q(t,x)$ in Eq. (3) by setting $a_3 = 0$. However, then the perturbation scheme loses first-order integrability, as the first-order equations cannot be satisfied for a general $u$ if only the coefficients $a_1$ and $a_2$ are available, and these two become indeterminable [6]. Whatever values they are assigned, part of the perturbation remains unaccounted for. The calculation forces one to incorporate this "left-over" part in the first-order contribution in Eq. (4), spoiling the integrability of the equation, and destroying the simple KdV-multiple-soliton structure of $u$.

This observation is compounded by the fact that, in general, except when $u$ is a single-soliton solution, Eq. (1) may lose asymptotic integrability from second order onwards [2-5]. (This actually happens in the cases of the shallow-water problem and the ion acoustic wave equations [10].) One cannot express the higher-order corrections. $u^{(n)}$, $n \geq 2$, as differential polynomials in $u$ and, *simultaneously*, derive a Normal Form that is updated and integrable through $O(\varepsilon^n)$ to govern the evolution of the zero-order approximation, $u$. The perturbation contains parts that constitute "obstacles to asymptotic integrability". The algebra fails to account for these parts, and one is forced to incorporate them in the Normal Form, thereby destroying its integrability. Obstacles to asymptotic integrability and the manner in which a second-order obstacle spoils the KdV structure of a two-soliton solution have been first exposed in [2-5]. The effect of the obstacle that is encountered if one insists on preserving the scattering picture in $O(\varepsilon)$ is the same. In summary, as long as the expressions for higher-order corrections are limited to differential polynomials in $u$, asymptotic integrability is not achievable beyond $O(\varepsilon)$. If, in addition, one insists on maintaining the $O(\varepsilon)$ scattering picture, then asymptotic integrability is lost in $O(\varepsilon)$ as well.

In the following, it will be shown that, in the multiple-soliton sector of solutions of the KdV equation, the loss of the scattering picture and of asymptotic integrability are both associated with the emergence of inelastically generated solitons and anti-solitons (negative-amplitude solitons) in the perturbation. These lead to the generation of soliton-anti-soliton waves in the $O(\varepsilon)$ contribution to the solution. The approach is based on the observation that none of the problems discussed above is encountered when the zero-order approximation, $u(t,x)$, is a single-soliton solution. The solution of Eq. (1) is constructed as a sum of a major, elastic, component and an $O(\varepsilon)$ inelastic component. The elastic component has the same structure in the single- and multiple-soliton cases. It preserves the scattering picture and enjoys all the characteristics of a solution of Eq. (1) when it is integrable equation (i.e., as if no obstacles to integrability exist) order-by-order. The inelastic component arises only in the multiple-soliton case. It is generated by perturbation terms that represent interaction between KdV solitons and inelastically generated non-KdV soliton-anti-soliton waves. These terms have the capacity to generate soliton-anti-soliton waves in the solution. This is demonstrated through the numerical analysis of the case of a two-soliton solution of the KdV equation. Far from the origin, the inelastic component tends to a soliton-anti-soliton pair. The latter have the same structure as the original KdV solitons. However, their amplitudes are different from those of the KdV solitons, and have opposite signs. The results are presented through $O(\varepsilon)$. The detailed analysis through $O(\varepsilon^3)$ will be presented in a future publication.

To obtain the elastic component through $O(\varepsilon)$, one chooses $a_3 = 0$. Noting that the problems discussed above do not arise in the single-soliton case, one assigns $a_1$ and $a_2$ their single-soliton values of Eq. (6), with $a_3 = 0$. With this choice, the unaccounted for part of the perturbation is:

$$10(\alpha_2 - \alpha_4)R_1(t,x) \qquad \left(R_1(t,x) = \partial_x R_0(t,x) \quad , \quad R_0(t,x) = u^3 - u_1^2 + u u_2\right) . \qquad (9)$$

By construction, this term vanishes identically when computed for a single-soliton solution. Hence, it reflects the net effect of the difference between the single-and multiple-soliton cases.

To preserve asymptotic integrability of the Normal Form, Eq. (4), one incorporates in $u^{(1)}$ a term, which is to account for the effect of the obstacle of Eq. (9). Eq. (3) is modified into

$$u^{(1)} = u_{el}^{(1)}(t,x) + \eta(t,x) \qquad \left(u_{el}^{(1)} = a_1 u_2 + a_2 u^2\right), \qquad (10)$$

Note that $\eta(t,x)$ cannot be written as a differential polynomial in $u$. Using Eq. (6) for $a_1$ and $a_2$, with $a_3 = 0$, $u$ obeys the Normal Form, Eq. (4), provided that $\eta(t,x)$ obeys the following equation:

$$\eta_t = 6\partial_x(u\eta) + \eta_{xxx} + 10(\alpha_2 - \alpha_4)R_1(t,x). \qquad (11)$$

Using Eqs. (3), (6) and (10), the solution of Eq. (1) can be written as:

$$w(t,x) = w_{el}(t,x) + \varepsilon\eta(t,x). \qquad (12)$$

The main, elastic, component is given through first order by

$$w_{el}(t,x) = u(t,x) + \varepsilon u_{el}^{(1)} + O(\varepsilon^2). \qquad (13)$$

It is the contribution of $u$, a multiple-KdV-soliton solution (with soliton velocities given by Eq. (5)), to which a first-order correction, $u_{el}^{(1)}$, that does not spoil the scattering picture, is added. $w_{el}$ exists in both the single-and multiple-soliton cases. The inelastic correction term, $\eta(t,x)$, exists only in the multiple-soliton case, and does not affect the elastic component.

At this point, revealing the structure of $R_1(t,x)$ and of $\eta(t,x)$ is required. $R_1(t,x)$ vanishes identically when $u$ is a single-soliton solution. Consequently, for a multiple-soliton solution it vanishes exponentially fast away from the origin, where each "leg" of $u$ asymptotes into a single-soliton [6]. A useful interpretation of $R_1(t,x)$ emerges if written as [13]:

$$R_1(t,x) = u^2 u_s \quad , \quad u_s = \partial_x \left( \frac{3u^2 + u_2}{u} \right) . \tag{14}$$

In the form of Eq. (14), $R_1(t,x)$ represents a coupling term involving the multiple-KdV-soliton wave, $u$, and $u_s$, which is a non-KdV multiple-soliton state. For example, in the two-soliton case, the asymptotic structure of $u_s$ away from the origin in the $x$-$t$ plane is:

$$u_s \xrightarrow[|t|\to\infty]{} \frac{a_1}{\left(\cosh\{K_1(x+V_1 t)\} + \xi_1\right)^2} + \frac{a_2}{\left(\cosh\{K_2(x+V_2 t)\} + \xi_2\right)^2} . \tag{15}$$

In Eq. (15), one has

$$a_1 = -\text{sgn}(t) \cdot 2|k_1 - k_2|(k_1 + k_2)^2 \quad a_2 = \text{sgn}(t) \cdot 2(k_1 + k_2)(k_1 - k_2)^2 , \tag{16}$$

and to lowest order in $\varepsilon$,

$$\begin{aligned} K_1 &= k_1 + k_2 \quad , \quad K_2 = |k_1 - k_2| \\ V_1 &= 4(k_1^2 + k_2^2 - k_1 k_2) \quad , \quad V_2 = 4(k_1^2 + k_2^2 + k_1 k_2) \end{aligned} , \tag{17}$$

Whereas $u$ is constructed by elastic combinations of the Jost functions of the inverse scattering problem associated with the KdV equation [14-20], $u_s$ is a soliton-anti-soliton pair, constructed from inelastic combinations of the same Jost functions. As the velocities of these new solitons are different from the velocities of the original KdV solitons, the characteristic lines of the two types of solitons overlap only near the origin. Thus, Eq. (14) provides additional insight regarding the exponential vanishing of $R_1(t,x)$ away from the origin.

The driving term in Eq. (11) may not satisfy the conditions of the Fredhom Alternative Theorem. Still, it does not generate unbounded contributions in $\eta(t,x)$. First, $R_1(t,x)$ is a complete differential (see Eq. (9)). Moreover, it is bounded and appreciable in the soliton-collision region (a finite neighborhood of the origin) and decays exponentially fast away from that region (so does $R_0(t,x)$). Hence, it cannot generate unbounded contributions along the characteristic lines of the KdV solitons in $u$. For instance, in the two-soliton case, it decays along the line of soliton no. 1 as:

$$R_1 \propto e^{-2|k_2(x+4k_2^2 t)|} \quad |k_1(x+4k_1^2 t)| \to C, \quad |k_2(x+4k_2^2 t)| \to \infty \quad . \tag{19}$$

The only potential pitfall is the possibility that $R_1(t,x)$ may generate secular terms away from the soliton characteristic lines, namely, in regions in the $x$-$t$ plane where $u$ itself falls off exponentially. These are the triangular sectors $\{|4 k_{(i-1)}^2 t| \ll |x| \ll |4 k_i^2 t|\}$ in the $x$-$t$ plane bounded by the characteristic lines of adjacent solitons. In these regions, Eq. (11) is reduced to:

$$\eta_t = \eta_{xxx} + 10(\alpha_2 - \alpha_4)R_1(t,x) \quad . \tag{20}$$

To resonate with the homogeneous part of Eq. (20), $R_1(t,x)$ must fall off as $exp[-k |(x + k^2 t)|]$ for some $k$. If it falls off at such a rate, then $\eta(t,x) = (x + a\,t)\,exp[-k\,|(x + k^2 t)\}]$ solves Eq. (20). Unless $a = k^2$, this solution is unbounded in parts of the $x$-$t$ plane. However, $R_1(t,x)$ cannot generate such solutions. For example, in the two-soliton case, it falls off in the triangular sectors as:

$$R_1 \propto e^{-2|k_1(x+4k_1^2 t)|-2|k_2(x+4k_2^2 t)|} \quad |k_1(x+4k_1^2 t)| \to \infty, \quad |k_2(x+4k_2^2 t)| \to \infty \tag{21}$$

Finally, in the case of a two-soliton $u$, the product in Eq. (14) changes sign as the origin is crossed, because of the sign change in $u_s$ (see Eqs. (15) and (16)). $R_1(t,x)$ shows two identical, peaks that have opposite signs and are slightly displaced in the *x-t* plane. In Fig. 1, $|R_1|$ is plotted, so that the graphics show both peaks. Because of their proximity, their integrated effect on $\eta(t,x)$ is small.

Let us now focus on $\eta(t,x)$. As $R_t(t,x)$ is a complete differential, and $\eta(t,x)$ is bounded, a conserved quantity emerges immediately, by integrating Eq. (11) over *x*:

$$\frac{d}{dt}\int_{-\infty}^{+\infty}\eta(t,x)dx = 0 \ . \tag{22}$$

This observation leads to a simplified analysis for

$$\omega(t,x) = \int_{-\infty}^{x}\eta(t,x)dx \ . \tag{23}$$

Using Eq. (9), Eq. (11) yields for $\omega(t,x)$ the following equation:

$$\omega_t = 6\omega_x u + \omega_{xxx} + 10(\alpha_2 - \alpha_4)R_0(t,x) \ . \tag{24}$$

Sufficiently far away from the origin, so that $R_0(t,x)$ can be neglected, Eq. (24) is reduced to:

$$\omega_t = 6\omega_x u + \omega_{xxx} \ . \tag{25}$$

A trivial solution of Eq. (25) is $\omega(t,x) = \alpha\, u(t,x)$ for any $\alpha$. As Eq. (25) is valid only in the domain in the *x-t* plane in which the KdV solitons are well-separated and asymptote each to a single-soliton solution, a more interesting possibility emerges when *u* is an *N*-soliton solution, namely, that $\omega(t,x)$ tends to $\alpha_i\, u(t,x)$ along the characteristic line of the $i^{th}$ KdV soliton ($1 \leq i \leq N$) with different coefficients assigned to different solitons. The driving term, $R_0(t,x)$, determines the coefficients $\alpha_i$, allowing for new types of solutions to emerge.

Eq. (24) has been solved numerically for $\omega(t,x)$ with $u(t,x)$ - a two-soliton solution, with the overall coefficient, $10(\alpha_2 - \alpha_4)$, replaced by 1. Fig. 2 shows $\omega(t,x)$ for a scattering process, i.e., with $\omega(t,x) \to 0$ for $t \to -\infty$. This corresponds to a solution of Eq. (1) that starts off with the elastic component only. The possibility just discussed is borne out. Within the numerical accuracy, the asymptotic behavior of $\omega(t,x)$ is described by a soliton-anti-soliton pair. These evolve along the characteristic lines of the KdV solitons contained in $u(t,x)$, and have exactly the same structure. Their amplitudes are determined by the "obstacle", $R_0(t,x)$. One soliton has a positive amplitude, the other - a negative amplitude. The asymptotic form of the pair is shown in Fig. 3. It is numerically indistinguishable from a sum of two single KdV solitons, the wave numbers, velocities and phase shifts of which are those of the original KdV solitons, but the amplitudes are borrowed from the numerical solution for $\omega(t,x)$. The conservation law, Eq. (22) is obeyed, because $\eta(t,x) = \partial_x \omega(t,x)$. Since $\omega(t,x)$ vanishes as $x \to \infty$, the conserved quantity is equal to zero.

The analysis through $O(\varepsilon^3)$ yields a zero-order approximation, $u$, which is governed by an integrable Normal Form, updated through third order. Hence, $u$ exhibits the same single- and multiple-soliton solutions as the unperturbed KdV equation. The velocity of each soliton is given by

$$v_i = 4 k_i^2 \left( 1 + \varepsilon \alpha_4 \left( 4 k_i^2 \right) + \varepsilon^2 \beta_8 \left( 4 k_i^2 \right)^2 + \varepsilon^3 \gamma_{14} \left( 4 k_i^2 \right)^3 + O(\varepsilon^4) \right) \; . \tag{26}$$

The elastic component of Eqs. (12) and (13) is updated as if no obstacles to integrability exist. Its structure in the multiple-soliton case is identical to that of the singe-soliton case, and is given by

$$w_{el}(t,x) = u(t,x) + \varepsilon u_{el}^{(1)}(t,x) + \varepsilon^2 u_{el}^{(2)}(t,x) + \varepsilon^3 u_{el}^{(3)}(t,x) + O(\varepsilon^4) \; , \tag{27}$$

Like $u_{el}^{(1)}(t,x)$ (Eq. (10)), the higher-order differential-polynomial corrections preserve the scattering picture exhibited by $u$ because they contain local terms only. For example, $u_{el}^{(2)}$ has the form

$$u^{(2)}(t,x) = b_1 u_4 + b_2 u u_2 + b_3 u_1^2 + b_4 u^3 , \qquad (28)$$

with known expressions for $b_k$, $1 \leq k \leq 4$.

The $O(\varepsilon^2)$ analysis leads to an updated Eq. (11) for the small inelastic component, $\eta(t,x)$:

$$\begin{aligned}\eta_t &= 6\partial_x(u\eta) + \eta_{xxx} + 10(\alpha_2 - \alpha_4) R_1(t,x) \\ &+ \varepsilon \begin{Bmatrix} 3\eta^2 + 10(2\alpha_3 - \alpha_2) u_x \eta_x \\ + 30\left(\alpha_2 u^2 + (\alpha_2 + \tfrac{1}{3}\alpha_3 - \tfrac{1}{2}(\alpha_1 + \alpha_4))u_{xx}\right)\eta + 10\alpha_2 u\eta_{xx} + \alpha_4 \eta_{xxxx} \end{Bmatrix} \\ &+ \varepsilon \partial_x R_2(t,x) + O(\varepsilon^2)\end{aligned} \qquad (29)$$

The $O(\varepsilon)$ driving term in Eq. (29) has the same properties as $R_1(t,x)$, the lowest-order driving term. It is a gradient of a differential polynomial in $u$, denoted here as $R_2$. It vanishes identically when computed for a single soliton solution. In the case of a multiple-soliton solution, it is localized in a small neighborhood of the origin in the $x$-$t$ plane, and vanishes exponentially fast away from the origin. It does not generate unbounded contributions in $\eta(t,x)$ for the very same reasons that the lowest-order driving term, $R_1$, does not. This term has been previously identified as the second-order obstacle to integrability [6, 13]. The results of the third-order analysis have a similar nature.

Finally, although the inelastic component destroys the scattering picture because $R_1(t,x)$ depends non-trivially on the wave numbers of both solitons (in particular, it is proportional to $(k_1 - k_2)^2$), the perturbation series for the full solution, preserves the multiple-soliton nature of the unperturbed solution. This is in agreement with [21], where the numerical analysis of the ion acoustic wave equations, demonstrated the robustness of multiple-soliton solutions against perturbations.

Acknowledgment Helpful discussions with G. Burde and L. Prigozhin are deeply acknowledged.

REFERENCES

1. Fokas, A. S. and Liu, Q. M., Phys. Rev. Lett. **77**, 2347=2351 (1996).
2. Kodama, Y., Phys. Lett. **112A**, 193-196 (1985).
3. Kodama, Y., Physica **16D**, 14-26 (1985).
4. Kodama, Y., Normal Form and Solitons, pp. 319-340 in *Topics in Soliton Theory and Exactly Solvable Nonlinear Equation*, ed. by M.J. Ablowitz et al. (World Scientific, Singapore, 1987).
5. Hiraoka, Y. & Kodama, Y., Normal Form and Solitons, Lecture notes, Euro Summer School 2001, The Isaac Newton Institute, Cambridge, August 15-25 (2002).
6. Veksler, A. and Zarmi, Y., Physica D, **217**, 77-87 (2006).
7. Korteweg, D. J. & De Vries, G., Phil. Mag., **39**, 422-443 (1895).
8. Washimi, H. and Taniuti, T, Phys. Rev. Lett., **17**, 996-998 (1966).
9. Karpman, V. I, *Non-Linear Waves in Dispersive Media* (Pergamon, Oxford, 1975).
10. Zarmi, Y. Physica D, submitted.
11. Kraenkel, R. A., Phys. Rev. E. **57**, 4775-4777 (1998).
12. Hirota, R., Phys. Rev. Lett., **27**, 1192-1194 (1971).
13. Veksler, A. and Zarmi, Y., Nonlinearity, **20**, 523-536 (2007).
14. Ablowitz, M. J., Kaup, D. J., Newell, A. C. and Segur, H., Stud. Appl. Math. **53**, 249-315 (1974).
15. Zakharov, V. E. & Manakov, S.V., Sov. Phys. JETP **44**, 106-112 (1976).
16. Ablowitz, M. J. & Segur, H., *Solitons and the Inverse Scattering Transforms* (SIAM, Philadelphia, 1981).
17. Novikov, S. P., Manakov, S. V., Pitaevskii, L. P. and Zakharov, V. E., *Theory of Solitons*, (Consultant Bureau, New York, 1984).
18. Newell, A. C., *Solitons in Mathematics and Physics*, (SIAM, Philadelphia, PA, 1985).
19. Olver, P. J., *Applications of Lie Groups to Differential Equations* (Springer-Verlag, New York, 1986).
20. Ablowitz, M. J. & Clarkson, P. A., *Solitons, Nonlinear Evolution Equations and Inverse Scattering* (Cambridge University Press, 1991).
21. Li, Y. and Sattinger, D. H., J. Math. Fluid Mech. **1**, 117-130 (1999).

FIGURE CAPTIONS

Fig. 1  Absolute value of $R_1$ (Eq. (9)) for two-soliton solution (Eq. (7), $k_1 = 0.1$, $k_2 = 0.2$).  Front peak is positive, rear peak is negative.

Fig. 2  Solution of Eq. (11) for $\omega(t,x)$ for two-soliton solution ($k_1 = 0.1$, $k_2 = 0.2$); $10(\alpha_2 - \alpha_4)$ replaced by +1.  Vanishing initial data.

Fig. 3  $\omega(t,x)$ for two-soliton solution ($k_1 = 0.3$, $k_2 = 0.2$), $t = 200$.

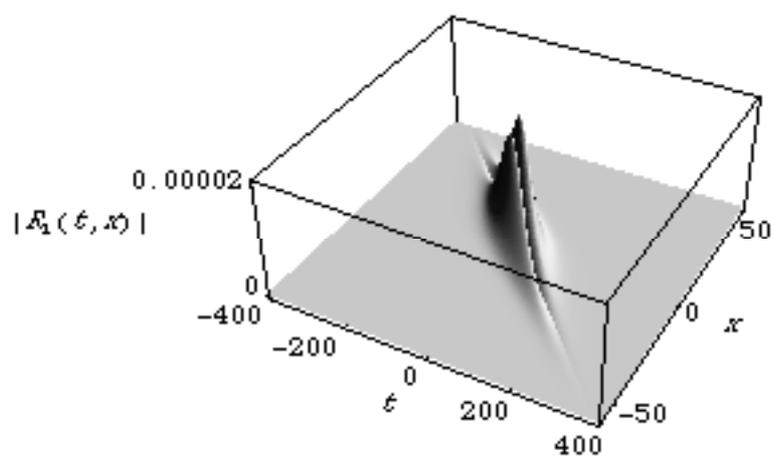

Fig. 1

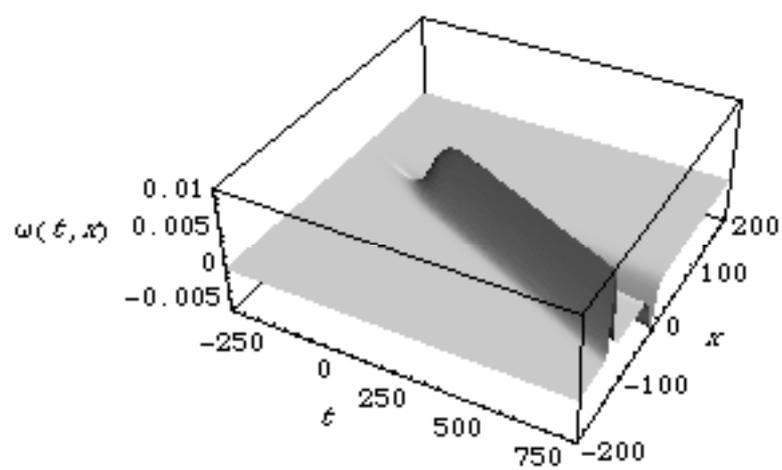

Fig. 2

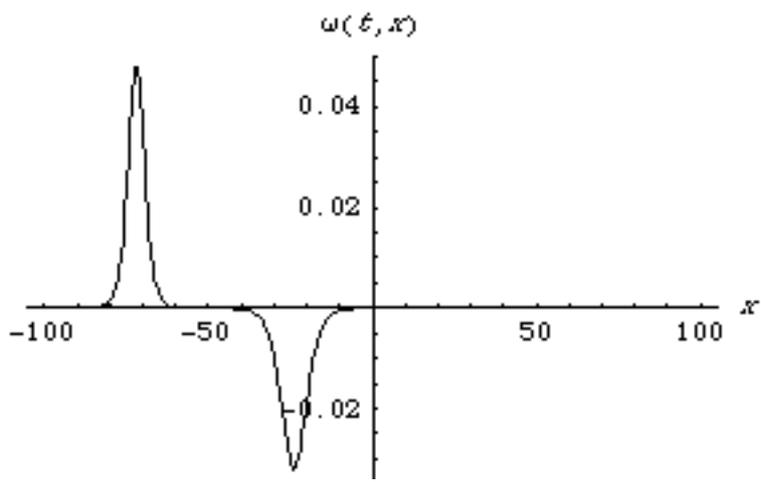

Fig. 3